\documentclass[intlimits,twoside,a4paper]{article}

\usepackage{amsmath,amssymb}
\usepackage{graphicx}

\usepackage[T2A]{fontenc}
\usepackage[cp1251]{inputenc}

\usepackage[eqsecnum]{cmpj2}

\issue{2016}{19}{1}{13606}
\doinumber{10.5488/CMP.19.13606}

\title[Self-sorting in two-dimensional assemblies]%
{Self-sorting in two-dimensional assemblies of simple chiral molecules}%
\author[A. Woszczyk, P. Szabelski]{A. Woszczyk, P. Szabelski\thanks{E-mail: szabla@vega.umcs.lublin.pl}}
\address{Department of Theoretical Chemistry, Maria-Curie Sklodowska University, \\ Pl.  M.C.~Sklodowskiej 3, 20-031 Lublin, Poland}

\date{Received September 29, 2015, in final form December 30, 2015}
\authorcopyright{A. Woszczyk, P. Szabelski, 2016}

\begin{document}

\maketitle

\begin{abstract}
Structural modification of adsorbed overlayers by means of external factors is an important objective in the fabrication of stimuli-responsive materials with adjustable physicochemical properties. In this contribution we present a coarse-grained Monte Carlo model of the confinement-induced chiral self-sorting of hockey stick-shaped enantiomers adsorbed on a triangular lattice. It is assumed that the adsorbed overlayer consists of ``normal'' molecules that are capable of adopting any of the six planar orientations imposed by the symmetry of the lattice and molecular directors having only one permanent orientation, that reflect the coupling of these species with an external directional field. Our investigations focus on the influence of the amount fraction of the molecular directors, temperature and surface coverage on the extent of the chiral segregation. The simulated results demonstrate that the molecular directors can have a significant effect on the ordering in enantiopure overlayers, while for the corresponding racemates their role is largely diminished. These findings can be helpful in designing strategies to improve methods of fabrication of homochiral surfaces and enantioselective adsorbents.
\keywords self-assembly, chiral molecules, Monte Carlo simulations, chiral resolution
\pacs 64.75.Yz, 68.43.Hn, 81.05.Xj, 02.70.Uu
\end{abstract}

\section{Introduction}

Achieving control over the structure of molecular systems, and tuning their properties using external inputs has been one the most intensively studied topics in material science during the recent years. A considerable interest in such adjustable molecular structures results mainly from the possibility of changing their physico-chemical characteristics in a reversible manner by means of noninvasive external factors including, for example, electric or magnetic fields. These directional inputs have been found to be capable of triggering the ordering in bulk systems comprising molecules and particles, which results in substantial changes in optical, magnetic or rheological properties~\cite{num1,num2,num3,num4}.

Another class of systems in which the structure formation can be guided by an external bias are the adsorbed overlayers comprising different, often functionalized, organic molecules. In this case, corrugation of the molecule-substrate interaction potential that is provided by atomic steps or grooves of metallic crystals has been often used to direct the self-assembly of molecular building blocks~\cite{num5,num6}. Moreover, this method, when combined with covalent linkage of the adsorbed molecules, has been effectively used to produce persistent macromolecular structures including wires, ribbons and networks~\cite{num7,num8}.

Regarding the directional external fields acting on 2D molecular assemblies, it has been demonstrated recently by Berg et al. that the structure of the adsorbed overlayer comprising prochiral functionalized biphenyl molecules can be imposed through the chiral imprinting process occurring at the liquid-solid interface~\cite{num3,num4}. In this approach, the prochiral molecules of the bulk liquid crystal (LC) phase were in direct contact with the graphite surface. The uniaxial ordering of the LC molecules that was induced by the directional magnetic field was the source of orientational confinement of the adsorbing molecules forming mirror-image homochiral domains on graphite. Systematic changes in the orientation of the applied field resulted in the variation of the relative population of the homochiral domains making it possible to create extended molecular assemblies with one handedness. The experiments by Berg and coworkers have clearly demonstrated that steering  the chiral organization of adsorbed molecules can be achieved using entirely achiral inputs such as directional magnetic fields. This idea seems particularly interesting from the perspective of chiral separations and fabrication of homochiral surfaces. For example, the field-induced orientational confinement of adsorbed enantiomers can be used to trigger their mixing or chiral resolution. These processes are relevant to the on-surface separation of enantiomers as well as to the creation of 2D molecular overlayers with adjustable adsorptive (enantioselective), optical and catalytic properties.

In order to explore the structure formation in adsorbed systems comprising chiral molecules, different theoretical approaches have been proposed including computer simulations and statistical mechanical calculations. In those studies, simplified molecular shapes, such as hard bent needles~\cite{num9,num10}, tetrominoes~\cite{num11}, tripods~\cite{num12,num13} or patchy disks~\cite{num14} have been usually employed, and the influence of anisotropic intermolecular interactions and molecular shape on the miscibility of the model enantiomers has been explored. Recently, more advanced approaches have been also proposed in which the adsorbed chiral mole\-cules were presented in a more detailed way, accounting for atomic charges, Lennard-Jones parameters of the composite atoms or functional groups as well as for corrugation of the \mbox{substrate~\cite{num15,num16,num17}}.

Even though the influence of intrinsic properties of chiral molecules on their performance in the 2D self-sorting process has been studied systematically, the effect of external anisotropic inputs has not yet been thoroughly explored. This refers especially to the role of the aforementioned orientational confinement of adsorbed enantiomers, which can be induced, for example, by external fields. Recently, using MC simulations, we have demonstrated the confinement-induced 2D chiral segregation of simple molecular tectons whose composite segments were allowed to occupy vertices of  a square lattice~\cite{num18,num19,num20}. These theoretical predictions suggest that the unidirectional positioning of surface enantiomers can be used to trigger the chiral resolution also on the surfaces having different symmetry. In this contribution we extend our MC investigations and explore the possibility of steering the chiral segregation that occurs on a triangular lattice. To that end, we consider simple hockey-stick-shaped molecules~\cite{num10} which are chiral, that is, they are capable of adopting mirror-image conformations when adsorbed. Moreover, in this work we study adsorbed overlayers in which only a part of the molecules is orientationally confined, playing the role of directors (sergeants) for the remaining molecules (soldiers).  The main objective of our simulations is to determine the effect of such parameters as the amount fraction of the director molecules, surface coverage and temperature on the extent of the ordering and chiral resolution in the systems investigated.

\vspace{-3mm}

\section{Simulation}

\begin{figure}[!t]
\centerline{
\includegraphics[width=0.62\textwidth]{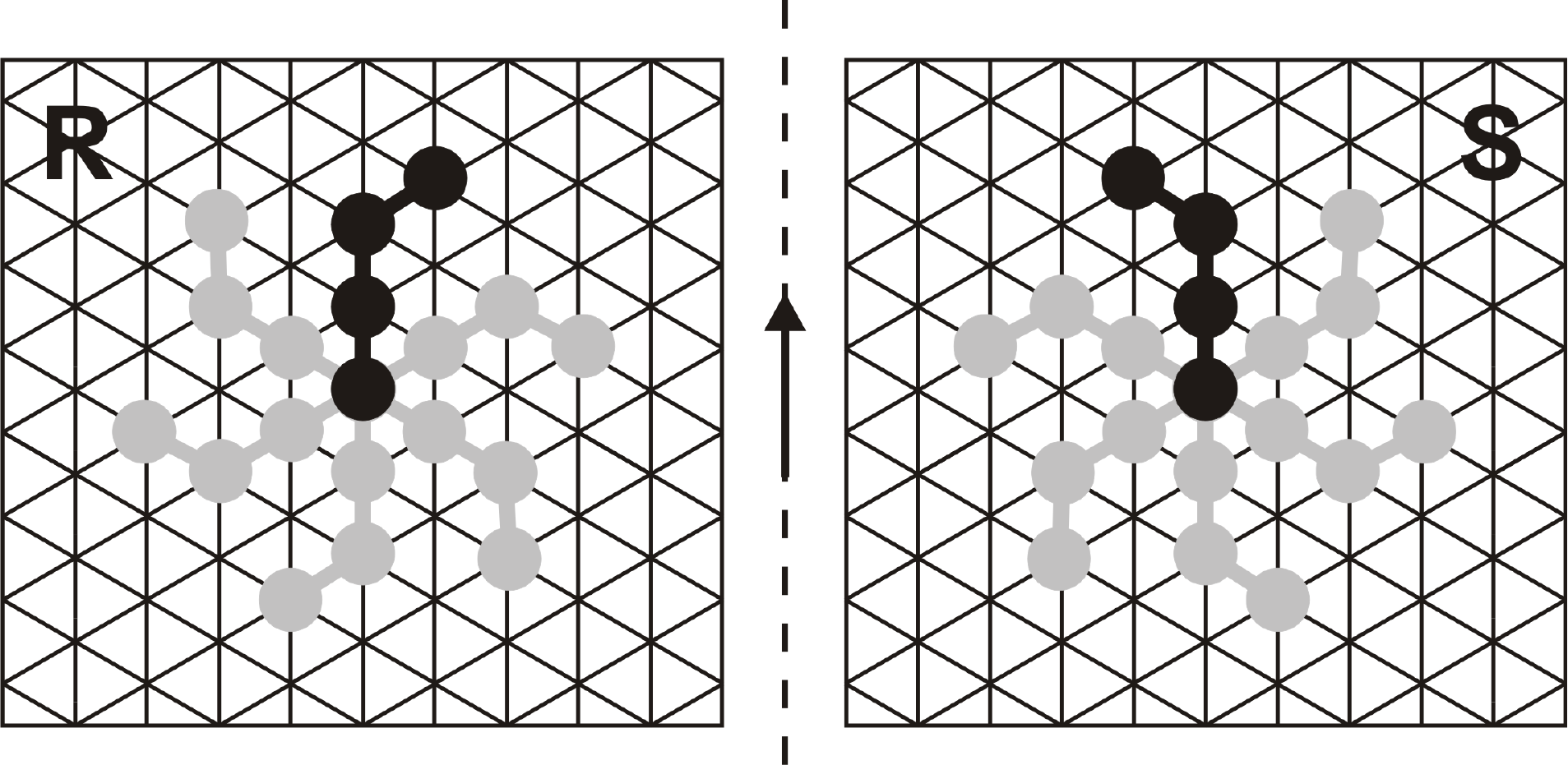}
}
\caption{Schematic structure of the four-membered enantiomers R and S adsorbed on a triangular lattice. The preferred orientation assumed for the director molecules is marked in black while the orientations accessible to the remaining molecules are shown in grey. The dashed line represents the mirror symmetry plane. The black arrow indicates the preferred orientation (long molecular arm) assumed in the model.}
\label{Fig1}
\end{figure}

The chiral molecules of our model were assumed to consist of four interconnected segments arranged in a hockey-stick shape shown schematically in figure~\ref{Fig1}. These molecules were allowed to adopt mirror-image configurations (R and S enantiomers) when adsorbed.
Each molecular segment of R and S occupied one site on a triangular lattice. The molecules were neither allowed to overlap nor to change chirality. Depending on the considered version of the model, the molecules could take any of the six planar orientations shown in the figure (isotropic) or the orientation of some of them was fixed to mimic the coupling of these species with the directional external field (see the molecules in black, unidirectional). The molecules with fixed orientation are further referred to as directors. In the simplified approach presented here, the director molecules have special properties which make them couple strongly with a directional external field while the rest of the molecules remain inert. In real situations, all of the molecules can be energetically coupled to the field and have orientationally dependent energies. In this case, the chiral segregation would depend on the individual energetic parameters which characterize each type of molecules. To minimize the number of parameters in our model, we considered an extreme case, in which the coupling energy of the director molecules is always large enough to orient these molecules parallel to the field. On the other hand, we also assumed that the coupling of the rest of the molecules is weak enough to be neglected. This approach permits to study the potential guiding role of the director molecules having a definite shape, without going into details of chemical composition of these species. Accordingly, the number of intermolecular interaction parameters of the model can be minimized. It was assumed that the interaction between adsorbed enantiomers can be described by an attractive short-range segment-segment interaction potential whose range was limited to the nearest neighbors on the lattice. The energy of elementary segment-segment interaction was characterized by the dimensionless parameter $\epsilon=-1$.

The MC simulations were performed on a 100 by 100 rhombic fragment of a triangular lattice imposing periodic boundary conditions in both planar directions. The calculations were carried out using the conventional MC method with Metropolis sampling~\cite{num18,num19,num20,num21,num22}, and they were additionally coupled with the cooling procedure in which the overlayer was equilibrated in a sequence of decreasing temperatures. The simulation protocol was organized as follows. In the first step $N$ molecules in total were randomly distributed on the surface. Depending on the assumed composition of the overlayer, homochiral (R) and racemic (R+S, 1:1) assemblies were modelled. Moreover, in each of these cases, the fraction of director molecules, $f$, was systematically changed, and for the racemate, the directors contributed equally to R and S. At the beginning of the MC run, the director molecules were randomly distributed on the surface together with the remaining molecules and their orientation was selected at random. This corresponds to a situation in which the director molecules are not initially  orientationally confined. Such a confinement, meaning a strong tendency to orient in one direction, can result from, for example, the coupling of the director molecules with external directional field parallel to the surface. During the simulation, when the confinement was imposed, the director molecules were allowed to translate but their orientation was changed (in case it was different from the original one) to the selected one shown in black in figure~\ref{Fig1}. To equilibrate the system, a series of random displacement moves was performed. In this procedure, the orientation of the director molecules was always fixed and the orientation of the remaining molecules was randomly changed by a multiple of 60 degrees. To accept or reject the move, the energies in the new $U_\text{n}$ and old $U_\text{o}$ positions were calculated for the selected molecule by summing up the segment-segment interactions with the neighboring molecules. Next, a random number $r \in (0,1)$ was generated and compared with the probability $p=\min[1,\exp(-\Delta U/k_\text{B}T)]$ where $\Delta U=U_\text{n}-U_\text{o}$ and $k_\text{B}$ and $T$ are the Boltzmann constant and temperature, respectively. If $r<p$, the move was accepted; otherwise it was rejected. In the simulations, we used dimensionless units, i.e., the energies are expressed in units of $\epsilon$ and temperature in units of $\lvert \epsilon \rvert/k$. In these calculations we used $2 \times 10^{8} \times N$   MC steps with one MC being a single attempt to move (and rotate if allowed) an adsorbed molecule. In such a MC run, the cooling procedure was realized within 200 temperature decrements of equal length; from $T=4.2$ down to $T=0.2$. Specifically, at each temperature, the system was equilibrated during $10^{6} \times N$ steps of which the last 10~\% were used for averaging. The adsorbed configuration equilibrated at one temperature was used for the subsequent lower temperature run. The cooling sequence was also used to calculate the corresponding specific heat capacities based on the fluctuations of the internal energy of the adsorbed overlayer~\cite{num23}. All  the results discussed in the manuscript are the averages over ten independent system replicas.

\section{Results and discussion}

\begin{figure}[!h]
\centerline{
\includegraphics[width=0.65\textwidth]{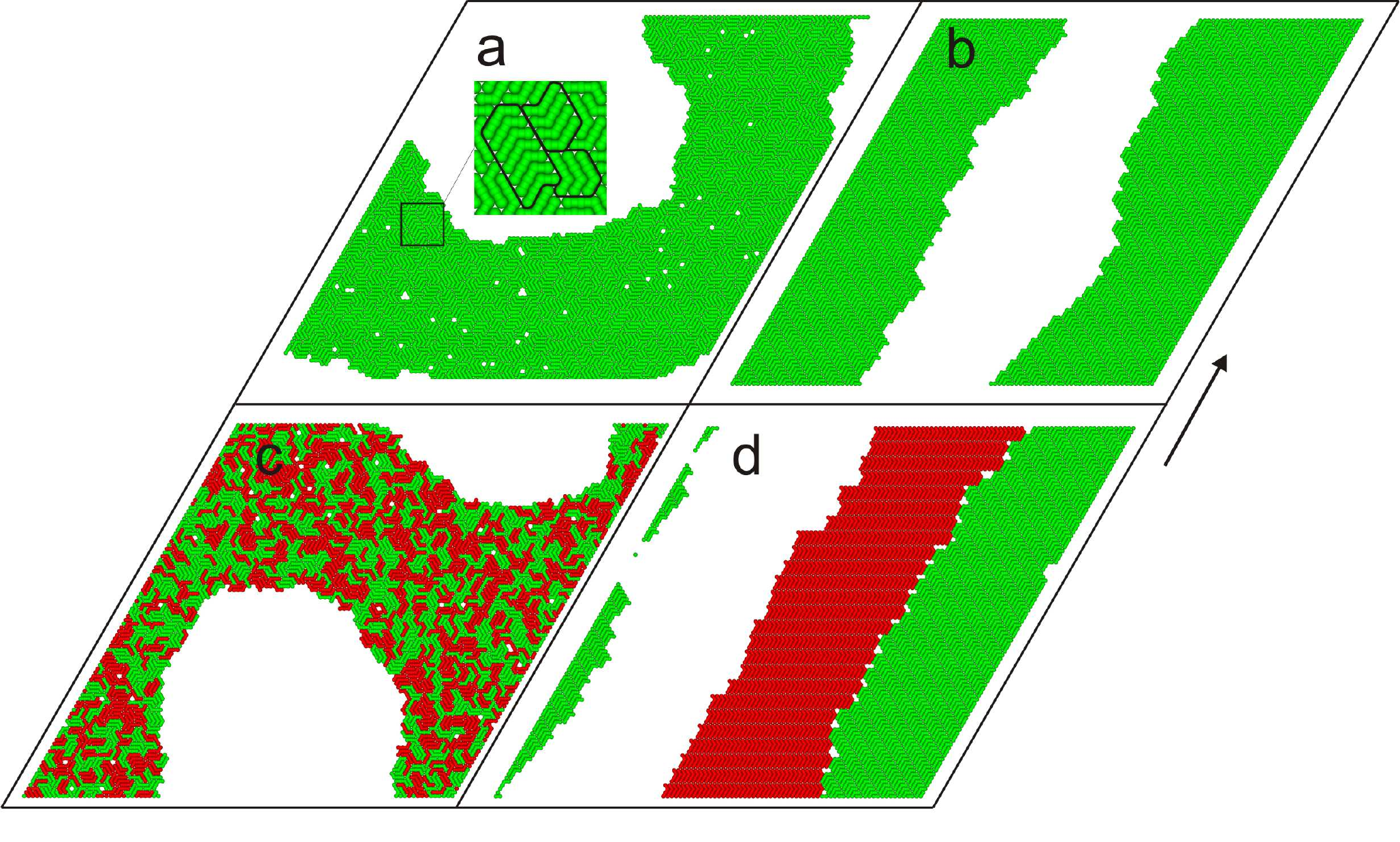}
}
\caption{(Color online) Snapshots of the overlayers comprising 1600 molecules adsorbed on a 100 by 100 triangular lattice; $T=0.2$. Panels (a) and (b) correspond to the enantiopure (R) systems while panels (c) and (d) show the results obtained for the corresponding racemates (800 R + 800 S). The effect of orientational confinement, indicated by the black arrow, on the structure formation is shown in the right-hand panels (b) and (d). The inset in panel (a) shows a magnified fragment of the overlayer in which clusters of uniformly aligned molecules can be noticed (encircled with black lines).}
\label{Fig2}
\end{figure}

To examine the effect of orientational confinement on the structure formation in our systems, we first performed simulations for the enantiopure overlayers. Top part of figure~\ref{Fig2} shows exemplary snapshots obtained for 1600 molecules R at $T=0.2$.
As it can be easily noticed in panel~(a), the rotational freedom of the adsorbed molecules is responsible for the formation of a compact disordered domain with sparse void defects. Although the obtained structure lacks a long-range order, some clustering of uniformly aligned molecules, shown in the inset, can be observed. Similar to puzzle tiles, the R clusters having different orientations are capable of packing densely and creating diversified local motifs. The tendency of R to form  the uniformly aligned local molecular structures becomes evident when the orientational confinement is switched on. In this case [panel~(b)], we can observe an extended lamellar pattern characterized by a rectangular $(1 \times 2 \sqrt{3})$ unit cell.

Regarding the racemic mixtures, in the case of unrestricted model, we can observe the formation of randomly mixed overlayers in which the enantiomers R and S exhibit some tendency to form homochiral clusters with uniform orientation [panel~(c)]. However, this trend is much less pronounced compared to the enantiopure overlayer and it results mainly from the increased number of local tightly-packed motifs which can be formed by R and S adsorbed together. The increased structural diversity of molecular packings in the racemic assembly prevents an efficient propagation of the enantiopure lamellar domains. These domains, as clearly seen in panel~(d), are formed when the orientational confinement is imposed on the enantiomers, leading to a complete demixing of R and S. The main source of the observed chiral resolution is the shape incompatibility of unlike enantiomers which are not capable of creating a dense mixed overlayer when uniformly oriented. The demixing is energetically favorable because it allows each molecule (R or S) to reach a maximum coordination (except for the phase boundary regions) characterized by the interaction energy equal to $9 \epsilon$.

\begin{figure}[!t]
\centerline{
\includegraphics[width=0.55\textwidth]{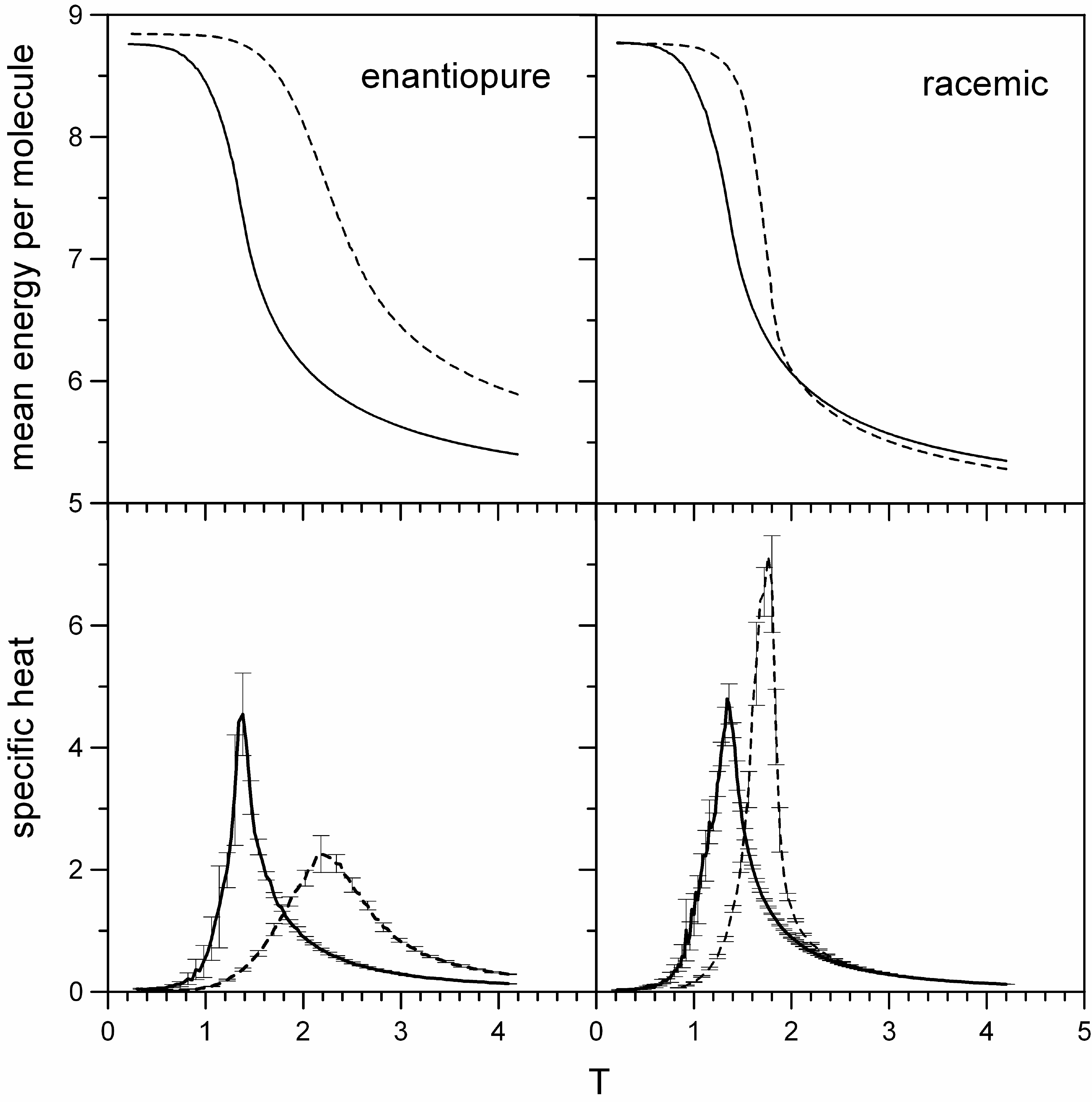}
}
\caption{Temperature dependence of the average potential energy of an adsorbed molecule (top) and specific heat capacity calculated for the isotropic (solid) and unidirectional (dashed) model, for the enantiopure and racemic overlayers comprising 1600 molecules. These results are the averages over ten independent system replicas.}
\label{Fig3}
\end{figure}

To quantify the differences between phase transformations occurring in the homochiral (a), (b) and racemic (c), (d) overlayers from figure~\ref{Fig2}, we calculated the mean potential energy of the adsorbed phase, $\langle U \rangle$ and the associated specific heat, $C_{v}$ as functions of temperature. These results are shown in figure~\ref{Fig3}.
As it is seen in the left-hand panels of figure~\ref{Fig3}, for the enantiopure systems we can observe a considerable shift in the position of the $\langle U \rangle$ and $C_{v}$ curves when the orientational confinement is switched on (dashed lines). Specifically, both $\langle U \rangle$ and $C_{v}$ move towards higher temperatures, from about 1.38 to 2.19, as estimated from the corresponding peak maxima. In this case, a strong tendency for the formation of an ordered overlayer that is inherent to the unidirectional model enhances the propagation of uniaxially aligned molecular pattern. Consequently, the transition is capable of occuring yet at higher temperatures. On the contrary, for the isotropic model, the increased orientational freedom of the adsorbed molecules lowers the temperature needed to reach the self-assembly. The results obtained for the enantiopure overlayers are qualitatively similar to those simulated for the corresponding racemates. However, for the racemic mixtures (right-hand panels), the analogous shifts in the position of the curves from the left-hand panels are much smaller. Here, the peak maximum in $C_{v}$ shifts from about 1.36 to 1.74 and, interestingly, the position of the transition corresponding to the isotropic model (solid line) is very close to that observed for the enantiopure system. The proximity of the transition points observed in $C_{v}$ for both types of isotropic models is likely to originate from the similar structure of the assemblies shown in the left-hand part of figure~\ref{Fig2}~(a), (c). For both these structures, we are dealing with randomly oriented, densely packed multimolecular clusters which fit in a way that a minimal number of void defects is created. In that sense, the presence of the other enantiomer (S) does not significantly change the mechanism of self-assembly. For the unidirectional model (dashed lines), however, the other enantiomer is responsible for an increased number of mixed molecular configurations (R-S) which can prevent somewhat the formation of separated homochiral domains. For this reason, the ordering transition shifts to lower temperatures.

\begin{figure}[tb]
\centerline{
\includegraphics[width=0.45\textwidth]{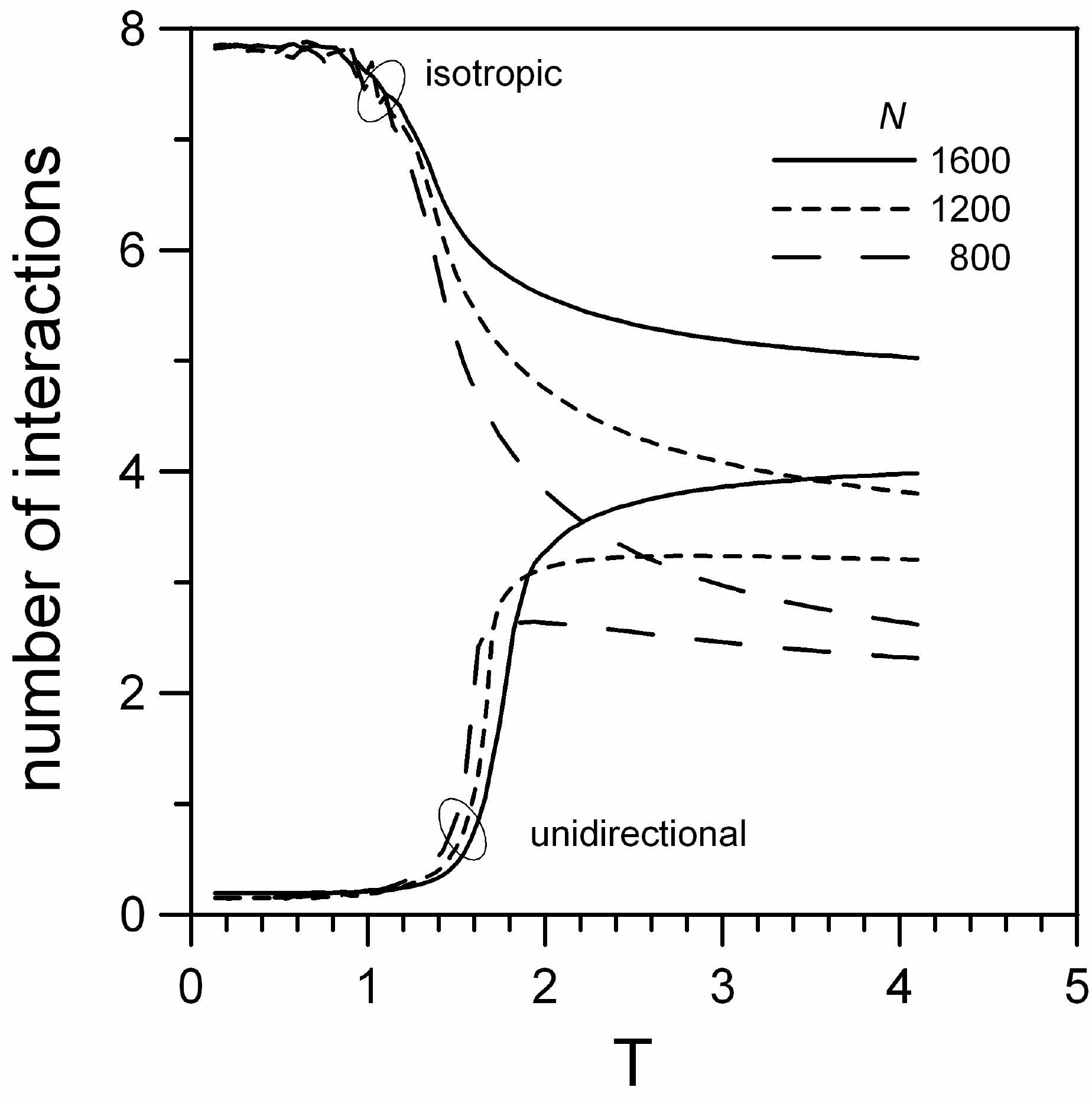}
}
\caption{The average number of heterogeneous interactions per molecule, $h$, as a function of temperature, calculated for the isotropic and unidirectional model, for a different number of adsorbed enantiomers (racemic overlayers). These results are the averages over ten independent system replicas.}
\label{Fig4}
\end{figure}

To trace the extent of chiral segregation in the racemic systems, we introduced the parameter $h$, which is equal to the average number of heterogeneous R-S intermolecular interactions per adsorbed enantiomer. In other words, this parameter provides the number of bonds that can be drawn between the composite segments of an R(S) molecule, and the neighboring segments belonging to S(R) molecules. The parameter $h$ characterizes the R-S interface length and, as it can be easily predicted, should be close to zero when the demixing occurs. On the other hand, for complete mixing of the enantiomers, $h$ should take the values close to 9, since in a compact racemic R-S domain, statistically half of 18 bonds of a molecule, say R (full coordination), are formed with S-segments. To visualize these dependencies in figure~\ref{Fig4} we plotted the parameter $h$ as a function of temperature, for both versions of the model, for different surface coverages.
From the results plotted in figure~\ref{Fig4}, it clearly follows that the phase transformations in the systems from figures~\ref{Fig2}~(c) and~\ref{Fig2}~(d) are manifested in the totally different shapes of the corresponding $h(T)$ curves. Namely, for the unidirectional model, we can observe a gradual decrease of $h$ with a decreasing temperature down to about zero, regardless of the surface coverage. This behavior is a clear indication of the chiral segregation and the residual value of $h \approx 0.2$ comes from the inter-domain R-S contacts in the demixed racemic systems. On the other hand, for the unrestricted model, $h$ increases with a decreasing temperature reaching a plateau at about 8. This value is consistent with the result of the simple mean field considerations made previously, meaning that the overlayer is randomly mixed with statistically half of the bonds  of a molecule being of the heterogeneous (R-S) type. Obviously, the obtained value $h \approx 8$ is lower than the ideal one (9) due to the clustering tendency of like molecules and due to the presence of peripheral molecules with fractional coordination. The relative position of the curves, in which the one corresponding to the mixing is always placed above the other curve, is a natural consequence of the increased number of bimolecular R-S configurations allowed in the unrestricted model. This can be seen even at high temperatures, close to 4.

\begin{figure}[!b]
\centerline{
\includegraphics[width=0.65\textwidth]{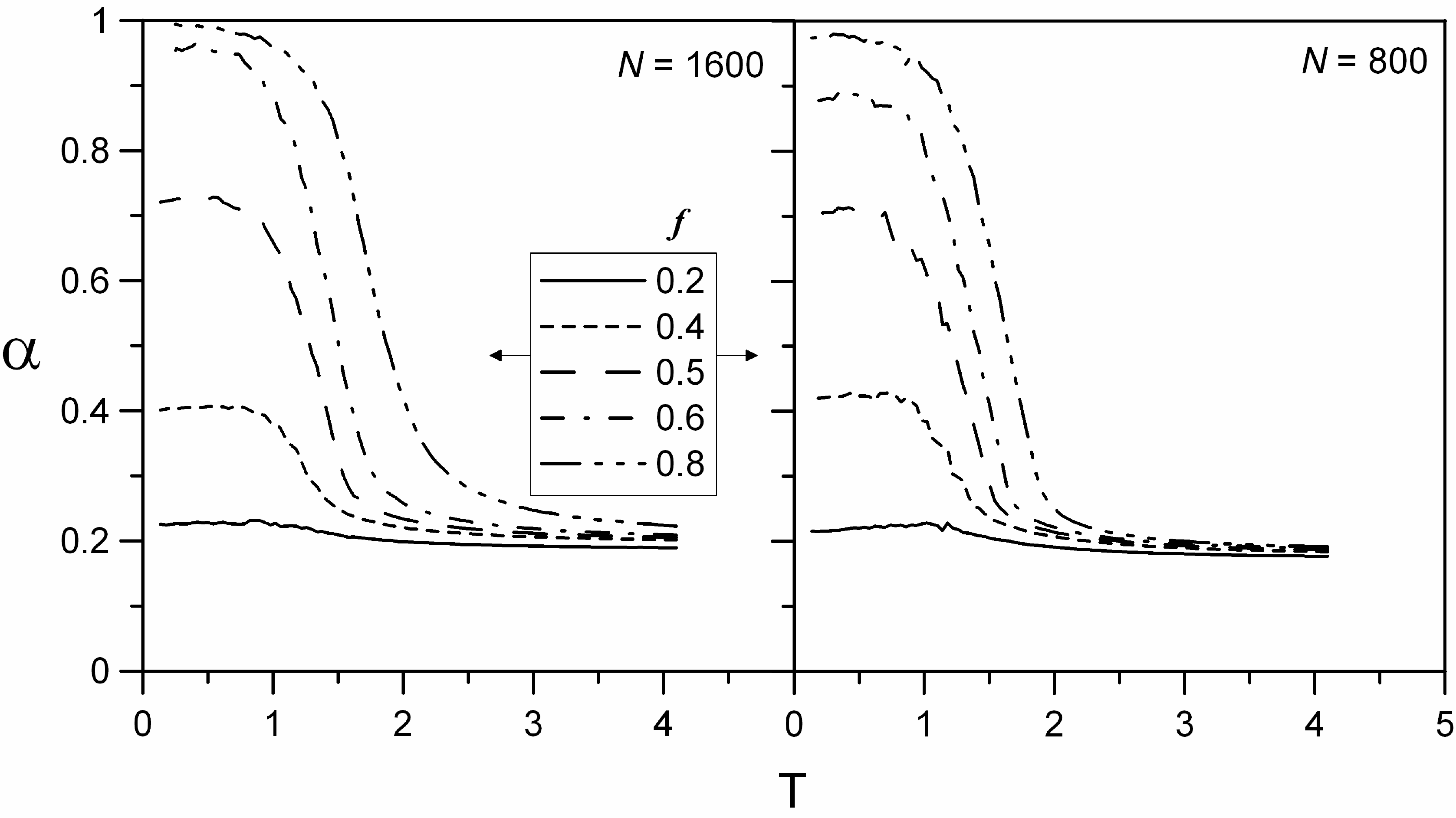}
}
\caption{Relative gain in the number of vertically oriented molecules, $\alpha$, as a function of temperature, calculated for different amount fractions of the director molecules. The results of the figure correspond to enantiopure overlayers with a different number of adsorbed molecules indicated in each panel. These results are the averages over ten independent system replicas.}
\label{Fig5}
\end{figure}

\begin{figure}[!h]
\centerline{
\includegraphics[width=0.7\textwidth]{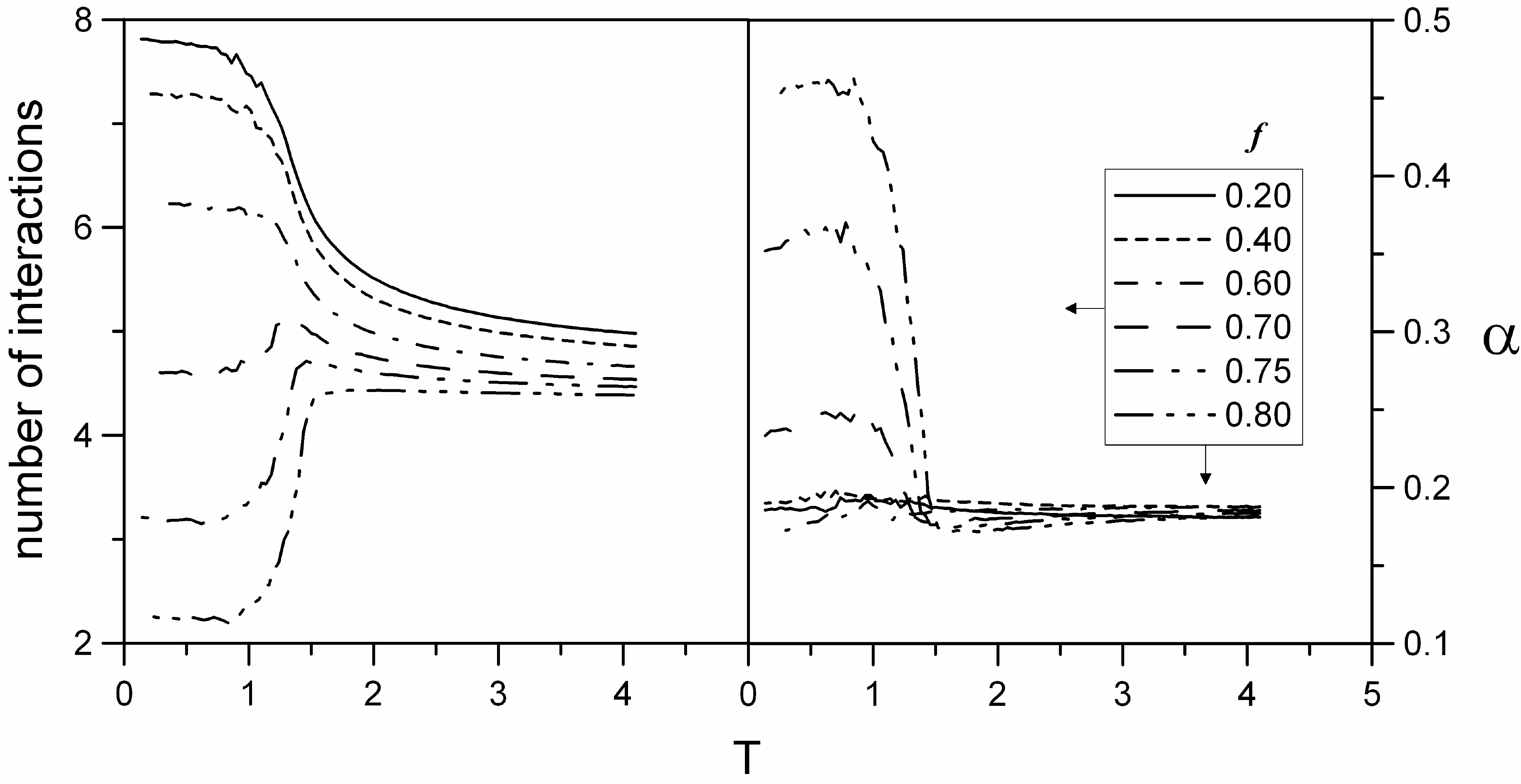}
}
\caption{The average number of heterogeneous interactions per molecule, $h$, and relative change in the number of molecules with upward orientation, $\alpha$, as functions of temperature calculated for the recamic overlayers for different amount fractions of the director molecules, $f$; $N=1600$. These results are averages over ten independent system replicas.}
\label{Fig6}
\end{figure}

\begin{figure}[!t]
\centerline{
\includegraphics[width=0.7\textwidth]{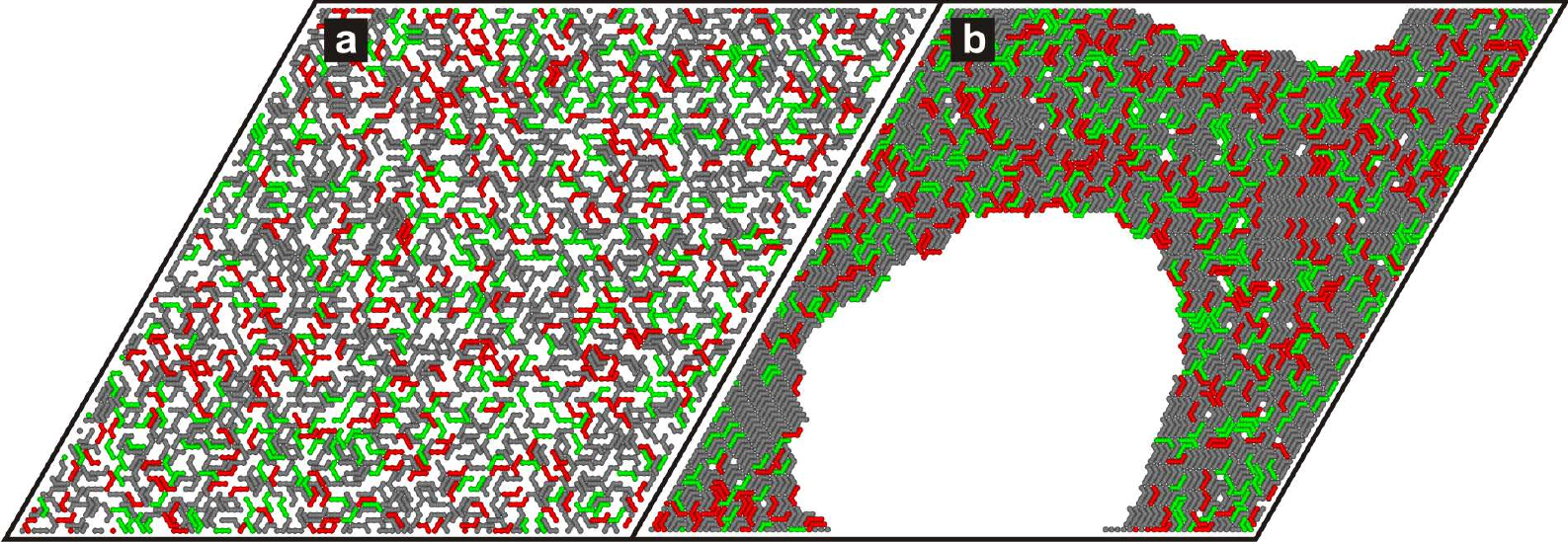}
}
\caption{(Color online) The initial (a) and final (b) configuration obtained for the racemic mixture comprising 1600 molecules (800 R + 800 S) of which 60~\% were the director molecules (dark grey, $f=0.6$).}
\label{Fig7}
\end{figure}

To explore the possibility of triggering the ordering transition using the director molecules we performed additional simulations in which the amount fraction, $f$, of these species was systematically chan\-ged. Figure~\ref{Fig5} presents the effect of the active dopant on the parameter $\alpha$, calculated for the enantiopure overlayers. This parameter is defined as follows:
\begin{equation}
\alpha=\frac{\hat{f}-f}{1-f}
\end{equation}
in which $\hat{f}$ is the actual fraction of molecules with an upward orientation (the same orientation as the one assumed for the directors, see figure~\ref{Fig1}). The parameter $\alpha$ measures the relative gain in the number of vertically oriented molecules that is induced by injecting the director molecules at fraction $f$.
From figure~\ref{Fig5}, it follows that in the enantiopure overlayers, below $T \approx 2$, the increase in the parameter $\alpha$ becomes noticeably large (> 0.5) when the fraction of the director molecules exceeds about one half. For example, for $f=0.6$, the fraction of vertically oriented molecules among all the remaining molecules (which are not the director ones) reaches over 90~\% highlighting  quite effective guided ordering mechanism. This also refers to the overlayers in which the coverage is lower (see the right-hand panel), and for which the obtained dependencies are very similar to those shown in the left-hand panel. To compare these results with the corresponding dependencies calculated for the racemates, in figure~\ref{Fig6} we plotted two additional sets of data. These, apart from the parameter $\alpha$, also include the average number of heterogeneous interactions per molecule, $h$.
The results of the left-hand panel of figure~\ref{Fig6} clearly demonstrate a diversified phase behaviour of the racemate upon adding the director molecules at a different fraction. Specifically, for $f$ lower than 0.6, we observe a distinct increasing trend in the average number of R-S heterogeneous interactions when the temperature drops. Here, we are dealing with an enhanced tendency to form a mixed R-S pattern. However, an increase of $f$ above 0.6 does not lead to a significant drop of $h$ (i.e., down to zero, see figure~\ref{Fig4}) at low temperatures where, for example, at $f=0.7$, $h$ is equal to about 4.7 which means an incomplete resolution of the enantiomers. To show this incomplete demixing occurring even at relatively high amounts of the director molecules, let us analyse the curves $\alpha$ from the right-hand panel of figure~\ref{Fig6}. In this case, we can observe that the relative increase in the fraction of vertically aligned molecules remains lower than 0.2 even for $f=0.6$. A further increase of $f$, however, does not lead to a significant growth (up to 1) of the parameter $\alpha$. On the contrary, even at a significant amount of the director molecules ($f=0.8$), the relative content of the new molecules with upward orientation is still lower than 50~\%. To illustrate how the director molecules contribute to the uniaxial ordering in figure~\ref{Fig7}, we showed an exemplary initial and final adsorbed configuration obtained for the system comprising 1600 molecules of which 60~\% were the director molecules (dark grey).
The results obtained for the racemic mixtures show that the role of the director molecules in the ordering in the corresponding 2D assemblies is much smaller than for the enantiopure systems. In other words, a much larger amount of the director molecules should be added to the racemate in order to induce its complete uniaxial ordering and the resulting chiral resolution. As mentioned previously, this effect originates mainly from the increased number of tightly packed molecular motifs which can be formed in the quasi-two component R-S system.

In the next figure~\ref{Fig8}, we presented the effect of the surface coverage (the number of adsorbed mole\-cules) on the extent of chiral resolution in the racemic overlayers.
As it follows from the left-hand panel, at a high director content, the curves calculated for the two lower coverages ($N=1200$ and $N=800$) exhibit a noticeable increase at moderate temperatures followed by steep drops after passing through the transition temperature (about 1.5). In these cases, the number of R-S interactions is naturally low at higher temperatures where we  deal with 2D gas phase. A further cooling of the adsorbed overlayer induces nucleation (increase in the number of R-S interactions), which is accompanied by the self-sorting effect leading to a decrease in $h$. Such a transition is not observed when $f$ is small (see the right-hand panel), and the nucleating aggregates preserve their mixed structure in the extended R-S domain formed at low $T$, which means that the director content is too low to induce the sorting process.

\begin{figure}[!b]
\centerline{
\includegraphics[width=0.7\textwidth]{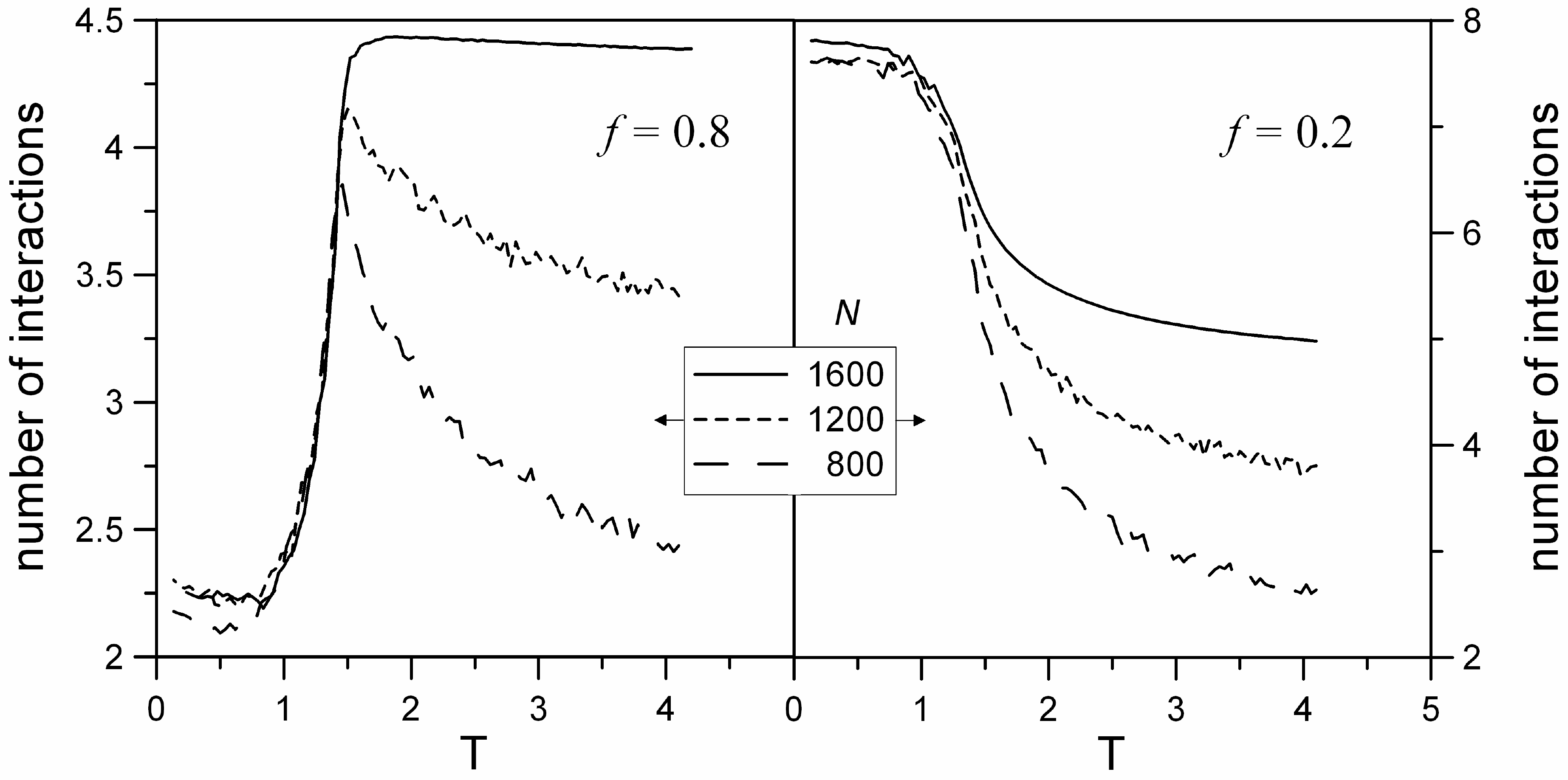}
}
\caption{Effect of surface coverage (number of adsorbed molecules, $N$) on the average number of heterogeneous interactions per molecule, $h$, calculated for racemic mixtures differing in the amount fraction of the director molecules, $f$ indicated in each panel. These results are averages over ten independent system replicas.}
\label{Fig8}
\end{figure}

The above results indicate that, regardless of the surface coverage, the self-sorting process is the most effective when the majority of the molecules of the racemate are forced to adopt one planar orientation. On the contrary, the role of the director molecules is found to be much larger for the enantiopure assemblies, which turned out to be more sensitive to the guiding influence of the active dopant.

\section{Conclusions}
In summary, using the coarse-grained MC modelling we have demonstrated that the orientational confinement of adsorbed chiral molecules can lead to their on-surface resolution producing homochiral ordered patterns. The calculations performed for the hockey-stick-shaped molecules, resembling the tetraphene, revealed the capability of this building block to form extended mirror-image domains upon 2D orientational confinement. It was found that the onset of phase transformation, in both enantiopure and racemic overlayers, shifts towards higher temperatures when the confinement is imposed on the molecules. However, this effect was more profound in the case of enantiopure systems highlighting the enhanced cooperativity of the like enantiomers in the structure formation. Separate simulations, in which only a part of the molecules was orientationally confined, showed that the chiral resolution (racemic) can be somewhat enhanced  using the guiding dopant molecules with a fixed in-plane orientation. However, in this case, nearly all of the adsorbed molecules, R and S, should be capable of aligning uniaxially and preserving this orientation.

The results of our theoretical investigations can help understand the role of a directional bias in the chiral resolution of enantiomers in adsorbed overlayers. These findings can be also useful in developing the methods of fabrication of homochiral surfaces, such as in designing strategies
to improve an adsorptive separation of chiral molecules.


\clearpage

\ukrainianpart

\title{Самосортування у двовимірних ансамблях простих хіральних молекул}%
\author{A. Вощик, П. Шабельський}
\address{Відділ теоретичної хімії, Університет Марії Кюрі-Складовської, 20-031 Люблін, Польща}

\makeukrtitle

\begin{abstract}
Структурна модифікація адсорбованих покриттів за рахунок зовнішніх факторів є важливим завданням при виготовленні стимуло-чутливих матеріалів з регульованими фізико-хімічними властивостями. У даній статті представлено огрублену модель індукованого просторовим обмеженням хірального самосортування енантіомерів із формою хокейної ключки, адсорбованих на трикутній гратці, яка досліджується методом Монте Карло.
Зроблено припущення, що адсорбоване покриття складається з
``нормальних'' молекул, які здатні прийняти будь-яку з шести планарних орієнтацій завдяки симетрії гратки і молекулярних директорів, що мають лише одну постійну орієнтацію, яка відображає їх зв'язок із зовнішнім полем певного напрямку.
Дане дослідження описує вплив фракції молекулярних директорів, температури та покриття поверхні на ступінь хіральної сегрегації.
Результати моделювання показують, що молекулярні директори  можуть мати значний вплив на впорядкування в енантічистих покриттях, в той час як для відповідних  расематів їх роль значно знижена. Дані дослідження можуть бути корисними для розробки стратегії вдосконалення способів виготовлення  гомохіральних поверхонь та енантіо-селективних адсорбентів.
\keywords самовпорядкування, хіральні молекули, моделювання Монте Карло, \\
хіральна роздільність
\end{abstract}

\end{document}